\documentstyle[multicol,aps,epsf]{revtex}

\begin{document}
\title{How Popular is Your Paper?  An Empirical Study of the Citation Distribution}
\author{S.~Redner}

\address{Center for Polymer Studies and Department of Physics, Boston
University, Boston, MA, 02215}
\maketitle
\begin{abstract}
  
  Numerical data for the distribution of citations are examined for: (i)
  papers published in 1981 in journals which are catalogued by the Institute
  for Scientific Information (783,339 papers) and (ii) 20 years of
  publications in Physical Review D, vols.\ 11-50 (24,296 papers).  A Zipf
  plot of the number of citations to a given paper versus its citation rank
  appears to be consistent with a power-law dependence for leading rank
  papers, with exponent close to $-1/2$.  This, in turn, suggests that the
  number of papers with $x$ citations, $N(x)$, has a large-$x$ power law
  decay $N(x)\sim x^{-\alpha}$, with $\alpha\approx 3$.

\end{abstract}
\begin{multicols}{2}

  In this letter, I consider a question which is of relevance to those for
  whom scientific publication is a primary means of scholarly communication.
  Namely, how often is a paper cited?  While the average or total number of
  citations are often quoted anecdotally and tabulations of highly-cited
  papers exist\cite{tab,spires}, the focus of this work is on the more
  fundamental {\em distribution of citations}, namely, the number of papers
  which have been cited a total of $x$ times, $N(x)$.  In spite of the fact
  that many academics are obliged to document their citations for merit-based
  considerations, there have been only a few scientific investigations on
  quantifying citations or related measures of scientific productivity.  In a
  1957 study based on the publication record of the scientific research staff
  at Brookhaven National Laboratory, Shockley\cite{sho} claimed that the
  scientific publication rate is described by a log-normal distribution.
  Much more recently, Laherrere and Sornette\cite{sor} have presented
  numerical evidence, based on data of the 1120 most-cited physicists from
  1981 through June 1997, that the citation distribution of individual
  authors has a stretched exponential form, $N(x)\propto
  \exp[-(x/x_0)^\beta]$ with $\beta\approx 0.3$.  Both papers give
  qualitative justifications for their assertions which are based on
  plausible general principles; however, these arguments do not provide
  specific numerical predictions.
  
  Here, the citation distribution of scientific publications based on two
  relatively large data sets is investigated\cite{data}.  One (ISI) is the
  citation distribution of 783,339 papers (with 6,716,198 citations)
  published in 1981 and cited between 1981 -- June 1997 that have been
  cataloged by the Institute for Scientific Information.  The second (PRD) is
  the citation distribution, as of June 1997, of the 24,296 papers cited at
  least once (with 351,872 citations) which were published in volumes 11
  through 50 of Physical Review D, 1975--1994.  Unlike Ref.~\cite{sor}, the
  focus here is on citations of publications rather than citations of
  specific authors.  A primary reason for this emphasis is that the
  publication citation count reflects on the publication itself, while the
  author citation count reflects ancillary features, such as the total number
  of author publications, the quality of each of these publications, and
  co-author attributes.  Additionally, only most-cited author data is
  currently available; this permits reconstruction of just the large-citation
  tail of the citation distribution.

\begin{figure}
\narrowtext
\epsfxsize=2.4in\epsfysize=2.4in
\hskip 0.3in\epsfbox{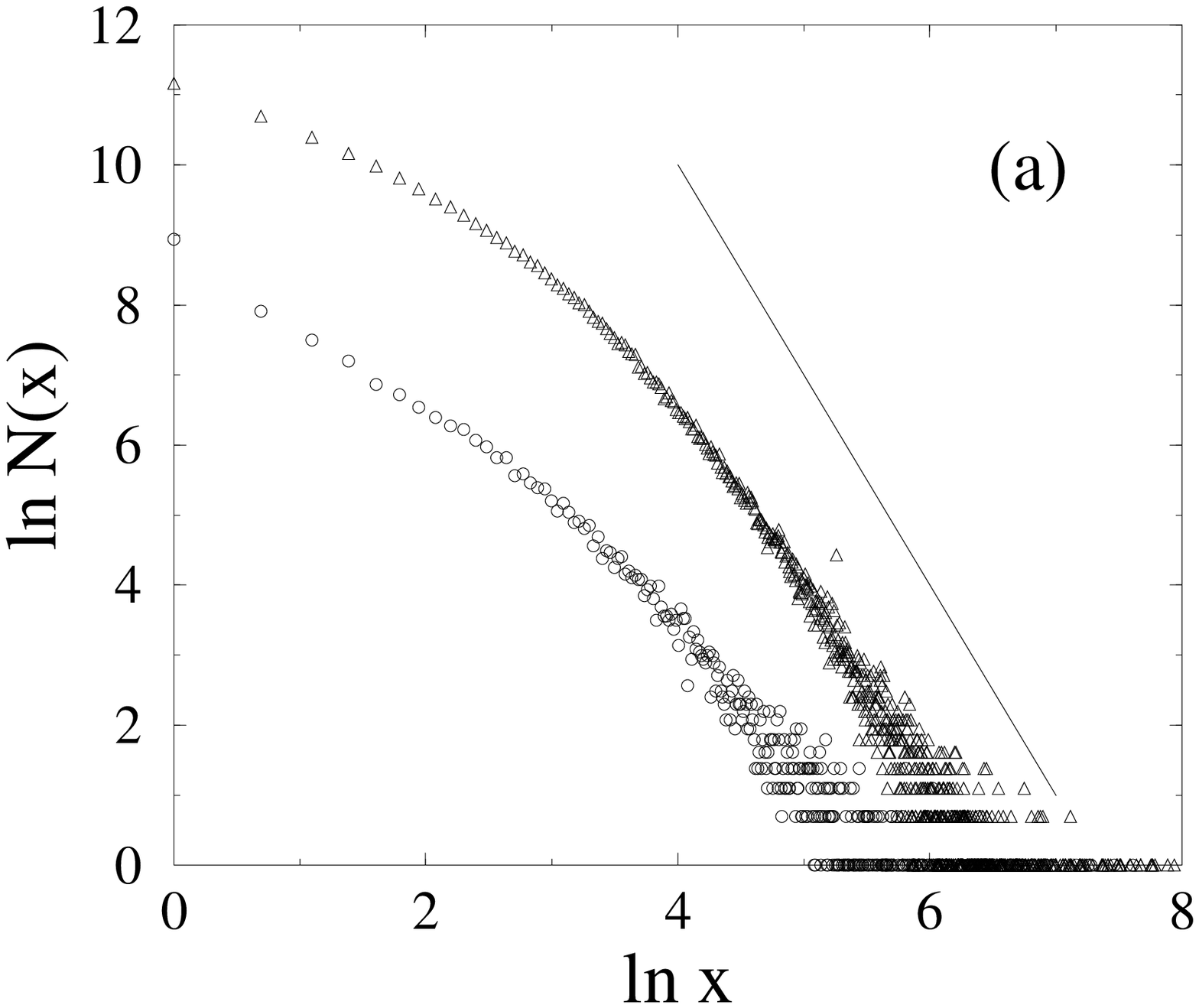}
\vskip 0.15in
\epsfxsize=2.4in\epsfysize=2.4in
\hskip 0.3in\epsfbox{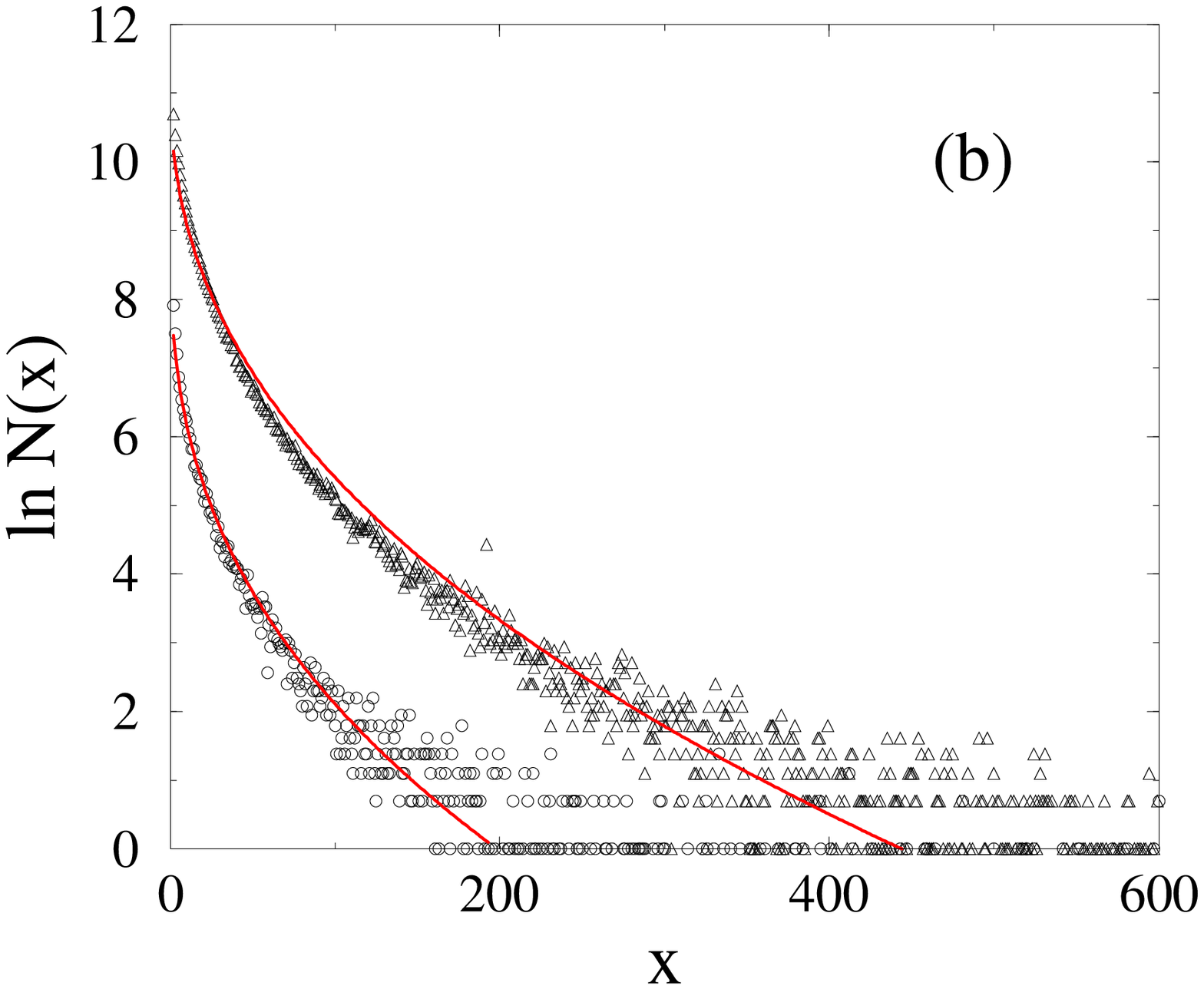}
\vskip 0.15in
\caption{(a) Citation distribution from the 783,339 papers in the ISI data
  set ($\Delta$) and the 24,296 papers in the PRD data set ($\circ$) on a
  double logarithmic scale.  For visual reference, a straight line of slope
  $-3$ is also shown.  (b) Same as (a), except on a semi-logarithmic scale.
  The solid curves are the best fits to the data for $x\leq 200$ (PRD)
  and $x\leq 500$ (ISI)
\label{fig1}}
\end{figure}

The main result of this study is that the asymptotic tail of the citation
distribution appears to be described by a power law, $N(x)\sim x^{-\alpha}$,
with $\alpha\approx 3$.  This conclusion is reached indirectly by means of a
Zipf plot (to be defined and discussed below), however, because Fig.~1
indicates that the citation distribution is not described by a single
function over the whole range of $x$.  

Since the distribution curves downward on a double logarithmic scale and
upward on a semi-logarithmic scale (Figs.~1(a) and (b) respectively), a
natural first hypothesis is that this distribution is a stretched
exponential, $N(x)\propto\exp[-(x/x_0)^\beta]$.  Visually, the numerical data
fit this form fairly well for $x\leq 200$ (PRD) and $x\leq 500$ (ISI) as
indicated in Fig.~1(b), with best fit values $\beta\approx 0.39$ (PRD) and
$\beta\approx 0.44$ (ISI).  However, the stretched exponential is unsuitable
to describe the large-$x$ data.  Here, data points are widely scattered,
reflecting the paucity of well-cited papers.  For example, in the ISI data,
only 64 out of 783,339 papers are cited more than 1000 times, 282 papers are
cited more than 500 times, and 2103 papers are cited more than 200 times,
with the most-cited paper having 8907 citations.  Such a sparsely populated
tail is not amenable to being directly fit by a smooth function.  (Amusingly
(or soberingly) 633,391 articles in the ISI set are cited 10 times or less
and 368,110 are uncited.)

Another test to determine the functional form of $N(x)$ is to compare
numerical values for the moments of the citation distribution
\begin{equation}
\label{mom}
\langle x^k\rangle = {\int x^k N(x)\, dx\over\int N(x)\, dx},
\end{equation}
with those obtained by assuming a given form for $N(x)$.  For example, if the
citation distribution is a stretched exponential, then the dimensionless
ratios ${\cal M}_k\equiv\langle x^k\rangle/\langle
x\rangle^k=\Gamma({k+1\over\beta}) \Gamma({1\over\beta})^{k-1}/
{\Gamma({2\over\beta})^k}$, where $\Gamma(x)$ is the gamma function.  Notice
that the scale factor $x_0$ in the exponential cancels.  For each $k$, an
estimate for $\beta$ can be inferred by matching the value of ${\cal M}_k$
obtained from the above gamma function formula with the corresponding
numerical data.  For both the ISI and PRD data, the corresponding estimates
for $\beta$ for $k=2, 3,\ldots, 6$ depend weakly but non-systematically on
$k$, and further do not match the values for $\beta$ obtained from a
least-squares fit to a stretched exponential (Fig.~1(b)).  Similarly, the
numerical data for $\langle x^k\rangle$ also do not match a power-law form
for the citation distribution, $N(x)\sim x^{-\alpha}$.  These results provide
evidence that the citation distribution is not described by a single function
over the entire range of citation count.

More fundamentally, it is natural to expect different underlying mechanisms
and different statistical features between minimally-cited and heavily-cited
papers.  The former are typically referenced by the author and close
associates, and such papers are typically forgotten a short time after
publication.  Evidence for such a short lifetime of minimally-cited papers
can be found, {\it e.g.}, by comparing the small-citation tail of $N(x)$ for
the first 4 years (1975-79) and the last 4 years (1990-1994) of the PRD data
set.  For $x\alt 200$, these data (appropriately normalized) and the complete
PRD data are virtually identical.  On the other hand, well-cited papers
become known through collective effects and their impact also extends over
long time periods.  This is reflected in the significant differences among
the large-citation tails of $N(x)$ for papers of different eras.

To help expose these differences in the citation distribution, it is useful
to construct a Zipf plot\cite{zipf}, in which the number of citations of the
$k^{\rm th}$ most-ranked paper out of an ensemble of $M$ papers is plotted
versus rank $k$ (Fig.~2).  By its very definition (see Eq.~(\ref{zipfe})),
the Zipf plot is closely related to the cumulative large-$x$ tail of the
citation distribution.  This plot is therefore well-suited for determining
the large-$x$ tail of the citation distribution.  The integral nature of the
Zipf plot also smooths the fluctuations in the high-citation tail and thus
facilitates quantitative analysis.

Given an ensemble of $M$ publications and the corresponding number of citations for
each of these papers in rank order, $Y_1\geq Y_2\geq \ldots
\geq Y_M$, then the number of citations of the $k^{\rm th}$ most-cited paper,
$Y_k$, may be estimated by the criterion\cite{extreme}
\begin{equation}
\label{zipfe}
\int_{Y_k}^\infty N(x)\,dx= k.
\end{equation}
This specifies that there are $k$ publications out of the ensemble of $M$
which are cited at least $Y_k$ times.  Eq.~(\ref{zipfe}) also represents a
one-to-one correspondence between the Zipf plot and the citation
distribution.  From the dependence of $Y_k$ on $k$ in a Zipf plot, one can
test whether it accords with a hypothesized form for $N(x)$.

In Fig.~2(a), a Zipf plot of the rank-ordered citation data is presented on a
double logarithmic scale for 4 data sets: (a) ISI data (top 200,000 papers
only), (b) complete PRD data (24,296 papers), (c) first 4 years of PRD data,
vols.\ 11-18 (5044 papers), and (d) last 4 years of PRD data, vols.\ 43-50
(5467 papers).  As alluded to previously, there is a considerable difference
between the first and last 4 years of the PRD data.  As might be anticipated,
the more recent highly-cited papers (up to approximately rank 700) are cited
less than papers in the earlier sub-data.  (There are two exceptions,
however.  These are the two top papers in the first 4 years which are cited
1741 and 1294 times, while in the last 4 years of data the two leading papers
are cited 2026 and 1420 times.)~ The larger citation count of heavily-cited
older papers reflects the obvious fact that popular but recent PRD papers are
still relatively early in their citation history.  This is in sharp contrast
to poorly-cited papers where there is little difference in the citation count
from the first 4 years and the last 4 years of the PRD data.

\begin{figure}
\narrowtext
\epsfxsize=2.4in\epsfysize=2.4in
\hskip 0.3in\epsfbox{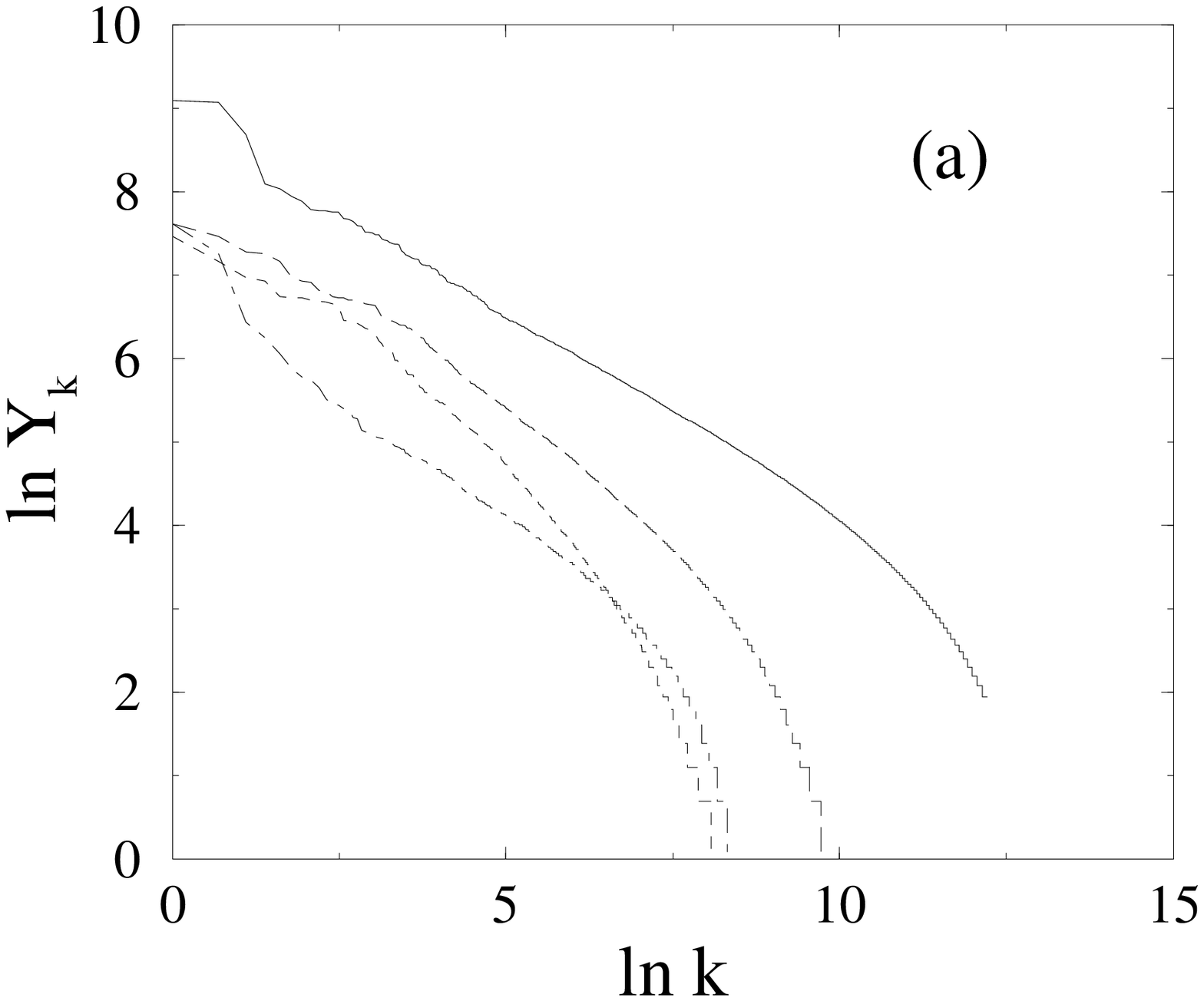}
\vskip 0.15in
\epsfxsize=2.4in\epsfysize=2.4in
\hskip 0.3in\epsfbox{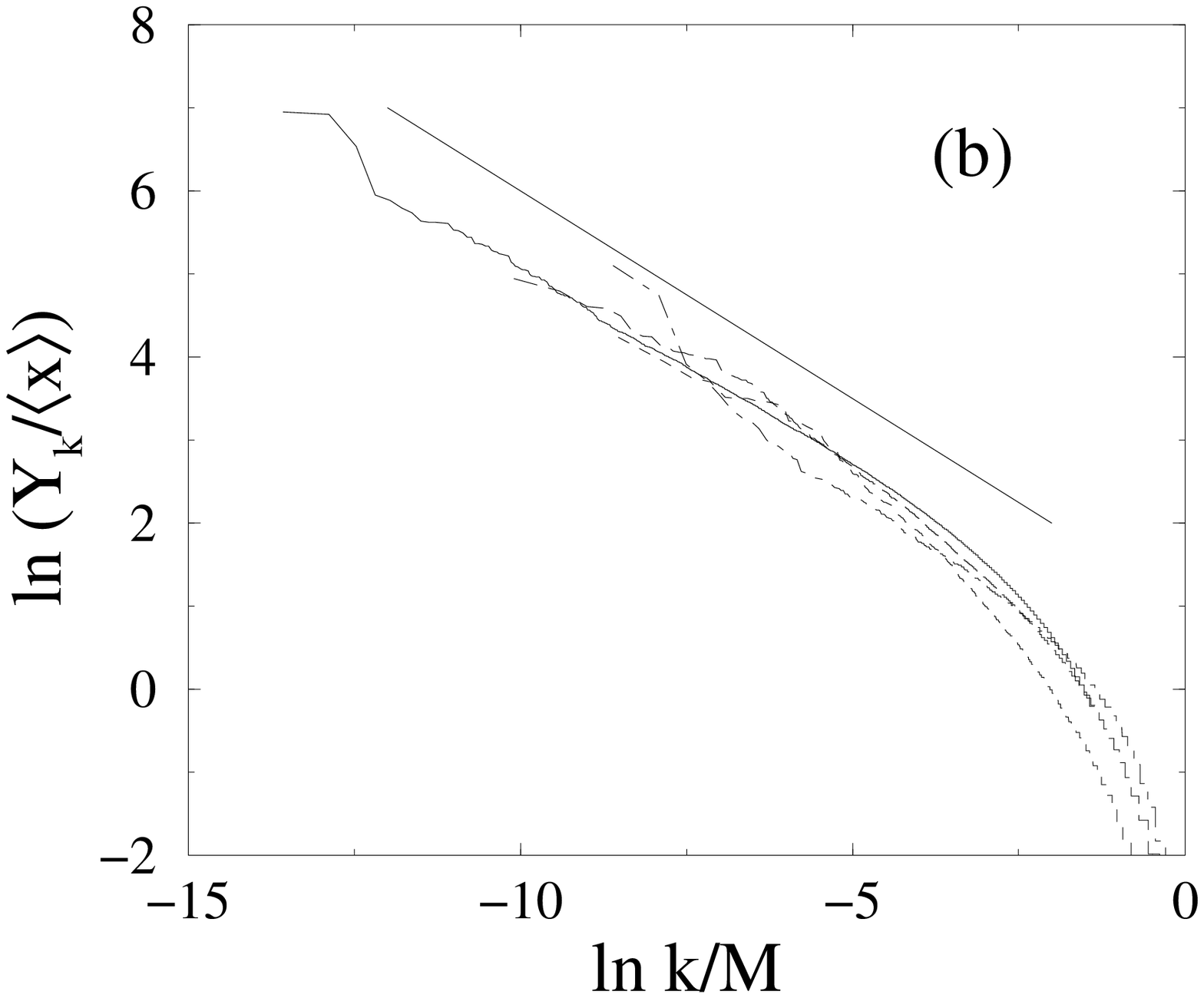}
\vskip 0.15in
\caption{(a) Zipf plot of the number of citations of the $k^{\rm th}$-ranked
  paper $Y_k$ versus rank $k$ on a double logarithmic scale.  (a) ISI data
  (---------), (b) PRD data (-- -- -- --), (c) vols.\ 11-18 of PRD (- - - -),
  (d) vols.\ 43-50 of PRD (-- $\cdot$ -- $\cdot$ -- $\cdot$ --).  (b) The
  data of (a) in scaled units.  For visual reference, a straight line of
  slope $-1/2$ is also shown.
\label{fig2}}
\end{figure}

To interpret the apparently unsystematic data in the Zipf plot of Fig.~2(a)
effectively, it is instructive to scale the data.  Since $k$ ranges between 1
and the number of publications in the ensemble, it is natural to define a
scaled relative rank $k/M_i$, where $M_i$ is the total number of papers in
each of the 4 data sets in Fig.~2, ($i=1$, 2, 3, or 4).  Similarly, for the
ordinate, it is useful to define a scaled citation count for the $k^{\rm th}$
most-cited paper by $Y_k/\langle x\rangle_i$, where $\langle x\rangle_i$ is
the average number of citations for all papers in the $i^{\rm th}$ data set.
As shown in Fig.~2(b), there is relatively good collapse of the 4 data sets
onto a single universal curve.  Notice also that the disparity in the two PRD
data subsets appears as a relatively small fluctuation about a mean value.
The data collapse also provides a strong clue about the location of the
asymptotic regime for citation data. 

Of particular relevance for the citation distribution, this scaling plot
indicates that the ISI data extends deeper than the PRD data into the
asymptotic tail of most heavily-cited papers.  This arises both from the fact
that the ISI data involves approximately 30 times more publications than the
PRD data and that the average number of citations to ISI papers is
approximately one-half that of PRD papers.  From the available data, the
highly-ranked ISI publications therefore provide the best representation of
the asymptotic tail of the citation distribution.  For ISI publications
between rank 1 (8904 citations) and 12,000 (approximately 85 citations), the
data is fairly linear and a least-squares data fit in this range yields an
exponent for the Zipf plot of Fig.~2(b) of approximately $-0.48$.  By
inverting Eq.~(\ref{zipfe}), this power law is equivalent to the distribution
of citations also having a power law form $N(x)\propto x^{-\alpha}$ for large
$x$, with $\alpha=1+1/.48\approx 3.08$.

This power-law behavior suggests a reconsideration of the citation
distribution in this range of $\geq$85 citations (Fig.~1(a)).  The curvature
in the data decreases significantly for $x\geq 85$ and it is not unreasonable
to attempt a power law fit, but with the additional caveat that the citation
data beyond approximately 500 citations is dominated by fluctuations.
Consequently there is subjectivity in specifying the range over which this
fit is performed.  Least-squares fits to the data within $50\leq x\leq 1000$
give exponent estimates in the range 2.6 -- 2.8, and a fit for $85\leq x\leq
500$, where the data are visually the most linear, both in the citation
distribution and in the Zipf plot, gives $\alpha\approx -2.7$.  The
correspondence between these fits and those in the Zipf plot therefore
suggest that the citation distribution may have a power-law tail, $N(x)\sim
x^{-\alpha}$, with exponent $\alpha$ close to 3.

\begin{table}
\caption{Annual citation data from PRD as of June 1997 including the first three 
moments of the citation distribution and the citation count of the most-cited paper.}
\begin{tabular}{|r|r|r|r|r|r|}
  Year & \# articles & $\langle x\rangle$ & $\langle x^2\rangle^{1/2}$ & $\langle x^3\rangle^{1/3}$ & $x_{\rm max}$ \\ \tableline
1975 & 1369 & 19.3 & 80.0 & 168.8 & 1294\\
1976 & 1085 & 17.8 & 71.8 & 178.5 & 1741\\
1977 & 1328 & 16.9 & 61.5 & 126.6 & 846\\
1978 & 1262 & 16.5 & 56.6 & 123.2 & 1066\\
1979 & 1229 & 17.8 & 57.5 & 114.7 & 907\\
1980 & 1114 & 18.5 & 61.9 & 126.2 & 912\\
1981 & 1107 & 17.0 & 62.5 & 148.8 & 1449\\
1982 & 1116 & 13.0 & 32.4 & 56.7 & 340\\
1983 & 1100 & 14.6 & 51.1 & 107.7 & 813\\
1984 & 1090 & 14.4 & 46.9 & 107.6 & 1004\\
1985 & 1094 & 13.8 & 34.6 & 66.7 & 579\\
1986 & 1222 & 11.8 & 24.7 & 38.9 & 215\\
1987 & 1275 & 12.7 & 35.6 & 79.5 & 772\\
1988 & 1124 & 11.4 & 23.6 & 38.8 & 244\\
1989 & 1153 & 12.0 & 29.8 & 62.6 & 600\\
1990 & 1161 & 13.0 & 29.7 & 57.5 & 515\\
1991 & 1083 & 13.2 & 34.5 & 74.9 & 622\\
1992 & 1388 & 13.3 & 47.3 & 130.0 & 1420\\
1993 & 1436 & 11.5 & 23.0 & 40.0 & 311\\
1994 & 1560 & 11.9 & 55.9 & 175.4 & 2026\\
\end{tabular}
\end{table}

Another important aspect of citation statistics which emerges from Fig.~2(a)
is its continuing temporal evolution.  This feature is nicely illustrated by
the annual citation statistics of PRD publications, where the average number
of citations $\langle x\rangle$ for articles published in a given year is
typically decreasing slowly with time (Table 1).  It is interesting that the
existence of a single exceptionally well-cited paper in a particular year has
an imperceptible effect on $\langle x\rangle$ but a much larger influence on
higher-order moments.  Notice also that the total number of citations to PRD
papers published in a given year, even as far back as 1975 (the first year
for which data is available), is slowly increasing.  Since papers from this
period which are still currently being cited are also likely to be highly
cited, this implies that the large-$x$ tail of the citation distribution has
not yet reached its final state.  Because of this continuing evolution of the
citation distribution, one cannot expect that the properties of the
high-citation tail of the citation distribution will be accurately determined
by direct analysis.

In summary, the citation distribution provides basic insights about the
relative popularity of scientific publications and provides a much more
complete measure of popularity than the average or total number of citations.
At a basic level, most publications are minimally recognized, with
$\approx47\%$ of the papers in the ISI data set uncited, more than 80\% cited
10 times or less, and $\approx .01\%$ cited more than 1000 times.  The
distribution of citations is a rapidly decreasing function of citation count
but does not appear to be described by a single function over the entire
range of this variable.  Although the available data is extensive, it still
appears insufficient to quantify the tail of the citation distribution
unambiguously by direct means.  However, a Zipf plot of the citation count of
a given paper versus its citation rank indicates a substantial range of power
law behavior with exponent close to $-1/2$.  This provides indirect evidence
that the citation distribution has a power law asymptotic tail, $N(x)\sim
x^{-\alpha}$ with $\alpha\approx 3$.  This differs from the conclusion of
Ref.~\cite{sor}, where the citation distribution of individual authors was
argued to have a stretched exponential tail.

Another important aspect of citations is that computerized data are
relatively recent, and the PRD data indicates that citation statistics from
1975 are still evolving.  Thus even the more extensive ISI data set is still
too recent to provide an accurate picture of the long-time and large-citation
tail of the citation distribution.  It should therefore be worthwhile to
study the properties of older citation data.  Alternatively, information of a
related genre, such as the distribution of sales for a particular class of
books, or ticket sales for movies and theaters may provide useful data for
studying citation-related statistics.

Finally, the citation distribution provides an appealing venue for
theoretical modeling.  There are several qualitative features about citations
which should be essential ingredients for a theory of their distribution.
Since almost all papers are gradually forgotten, the probability that a given
paper is cited should decrease in time with a relatively short memory.
Conversely, a paper which is in the process of becoming recognized gains
increasing attention through citations.  This suggests that the probability
of a paper being cited at a given time should be an increasing function of
the relative number of citations to that paper from an earlier time period.
Work is in progress to construct a model for the citation distribution which
is based on these considerations.

I thank P. Krapivsky, F. Leyvraz, and D. Sornette for helpful discussions, P.
Krapivsky for critical suggestions on the manuscript, and D. Sornette for
providing the author-based ISI citation data and related pertinent
information.  I also thank H. Galic from the SPIRES High-Energy Physics
Databases at SLAC, and D. Pendlebury and H. Small from the Institute for
Scientific Information for providing citation data and helpful advice.  I
gratefully acknowledge NSF grant DMR-9632059 and ARO grant DAAH04-96-1-0114
for partial financial support.

\end{multicols}


\begin{thebibliography}{99}
  
\bibitem{tab} See {\it e.g.}, {\it Science Citation Index Journal Citation
    Reports} (Institute for Scientific Information, Philadelphia) for annual
  lists of top-cited journals and articles (web site:
  http://www.isinet.com/welcome.html).  
  
\bibitem{spires} For example, current lists of top-cited articles in
  high-energy physics are maintained by the SPIRES High-Energy Physics
  Database at SLAC (web site http://www.slac.stanford.edu/find/top40.html).

\bibitem{sho} W. Shockley, Proc.\ IRE {\bf 45}, 279 (1957).

\bibitem{sor} J. Laherrere and D. Sornette, cond-mat/9801293.
  
\bibitem{data} The PRD data was provided by H. Galic from the SPIRES
  Database.  The ISI data was provided by D. Pendlebury and H. Small of the
  Institute for Scientific Information.  These two data sets and related
  citation data are available from my web site
  http://physics.bu.edu/$\sim$redner.

\bibitem{zipf} G. K. Zipf, {\it Human Behavior and the Principle of Least
    Effort} (Addison-Wesley, Cambridge, 1949).

\bibitem{extreme} This is a basic exercise in extreme value statistics.  See
  {\it e.g.}, J. Galambos, {\sl The Asymptotic Theory of Extreme Order
    Statistics}, (J. Wiley \& Sons, New York, 1978).


\end{thebibliography}
\end{document}